\documentclass[12pt]{article}

\usepackage{scicite}
\usepackage{times}
\usepackage{epsfig,amsmath}
\usepackage{subfigure}
\usepackage{graphicx}
\usepackage{color}

\newenvironment{sciabstract}{%
\begin{quote} \bf}
{\end{quote}}

\newcounter{lastnote}

\newcommand{\la}{\langle}
\newcommand{\ra}{\rangle}

\newcommand{\da}{\dagger}

\newcommand{\si}{\sigma}

\newcommand{\om}{\omega}

\newcommand{\non}{\nonumber}
\newcommand{\pa}{\partial}

\def\pra#1{{ Phys.\ Rev. A\/} {\bf#1}}
\def\prb#1{{ Phys.\ Rev. B\/} {\bf#1}}

\def\prl#1{{ Phys.\ Rev.\ Lett.} {\bf#1}}
\def\pr#1{{ Phys.\ Rev.} {\bf#1}}
\def\sci#1{{ Science} {\bf#1}}
\def\annph#1{{ Ann.\ Phys.} {\bf #1}}

\def\rmp#1{{ Rev. \ Mod. \ Phys.} {\bf#1}}
\def\nat#1{{ Nature} {\bf#1}}

\def\njp#1{{ New. J. \ Phys.} {\bf#1}}

\title{A quantum phase transition induced by a microscopic boundary condition}

\author
{Jun Jing,$^{1,2}$ Mike Guidry,$^{3}$ Lian-Ao Wu$^{2,4\ast}$\\
\\
\normalsize{$^{1}$Department of Physics, Zhejiang University}\\
\normalsize{Hangzhou 310027, Zhejiang, China}\\
\normalsize{$^{2}$Department of Theoretical Physics and History of Science,}\\
\normalsize{The Basque Country University (EHU/UPV), PO Box 644, 48080 Bilbao, Spain}\\
\normalsize{$^{3}$Department of Physics and Astronomy, University of Tennessee}\\
\normalsize{Knoxville, Tennessee 37996, USA}\\
\normalsize{$^{4}$Ikerbasque, Basque Foundation for Science, 48011 Bilbao, Spain} \\
\normalsize{$^\ast$To whom correspondence should be addressed; E-mail: lianao.wu@ehu.es}
}

\date{}

\begin{document}


\baselineskip24pt


\maketitle


\begin{sciabstract}
Quantum phase transitions are sudden changes in the ground-state wavefunction of a many-body system that can occur as a control parameter such as a concentration or a field strength is varied. They are driven purely by the competition between quantum fluctuations and mutual interactions among constituents of the system, not by thermal fluctuations; hence they can occur even at zero temperature. Examples of quantum phase transitions in many-body physics may be found in systems ranging from high-temperature superconductors to topological insulators. A quantum phase transition usually can be characterized by nonanalyticity/discontinuity in certain order parameters or divergence of the ground state energy eigenvalue and/or its derivatives with respect to certain physical quantities. Here in a circular one-dimensional spin model with Heisenberg XY interaction and no magnetic field, we observe critical phenomena for the $n_0=1/N\rightarrow0$ Mott insulator caused by a qualitative change of the boundary condition. We demonstrate in the vicinity of the transition point a sudden change in ground-state properties accompanied by an avoided level-crossing between the ground and the first excited states. Notably, our result links conventional quantum phase transitions to microscopic boundary conditions, with significant implications for quantum information, quantum control, and quantum computing.
\end{sciabstract}


\paragraph*{Introduction.}
Phase transitions are common macroscopic/bulk phenomena in both classical and quantum mechanical regimes~\cite{Sachdev,QPT2,QPT3,QPT4}. They can be organized into three basic categories: (i)~discontinuous or first-order phase transitions~\cite{1stQPT}, (ii)~continuous or second-order phase transitions~\cite{2ndQPT}, and (iii)~topological Kosterlitz--Thouless transitions between bound vortex-antivortex pairs at low temperatures and unpaired vortices and anti-vortices at high temperature~\cite{KT1,KT2,KT3,TI}. The ground state of a quantum system is the resource that enables quantum adiabatic passage~\cite{AP1,AP2,AP3} and quantum annealing~\cite{QA1,QA2,QA3}, which are of great relevance to  quantum information theory, quantum computation~\cite{Nielsen}, and quantum control~\cite{Wiseman}. It is then important to understand the general conditions under which a system can experience a quantum phase transition that alters fundamentally the structure of the ground state. The wavefunction is in general a more sensitive probe than the energy of the quantum structure for the ground state. It is expected that the qualitative changes of the ground-state wavefunction at a quantum phase transitions could have a dramatic effect on its role in quantum information processing.

There is no global theory for quantum phase transitions and for the behavior near quantum critical points for many-body and many-degree-of-freedom systems, but some basic features are reasonably well understood. As summarized by Sachdev~\cite{Sachdev}, a first-order quantum phase transition implies a sudden change of the ground-state wavefunction, accompanied by a singularity in the ground-state energy and a level-crossing or an avoided crossing between the two lowest states. A second-order or continuous quantum phase transition exhibits a singularity in the derivatives (in any order) of the ground-state energy with respect to some order parameter at the critical point, which is associated with the scaling behavior such as bipartite entanglement of a many-body system~\cite{Osterloh,WuPRL,WuPRA}.

Typically, critical behavior in a finite-size system becomes progressively sharper as the size tends to the infinite limit. One can check generically that the energy gap between the ground and the first excited states vanishes as a power-law function of the difference of some characteristic energy scale from its value at the critical point. In most cases known to date these energy scales are determined by a symmetry broken through local fluctuations. The corresponding quantum phase transition is tuned by varying a control parameter that either controls the strength of an applied perturbation (For example, an external transverse magnetic field disrupts the magnetic order of a quantum Ising model~\cite{Osterloh,Ising2,Ising3}), applied pressure, or the level of doping with electron donors or acceptors. However, quantum phase transitions could also be induced by novel effects, such as boundary conditions that modify the topological structure, and could occur in systems with a finite number of sites but with an infinite number of degrees of freedom~\cite{JC}. Study of such effects could be fertile ground for discovering new classes of quantum phase transitions that might be relevant for quantum information processing. This paper reports on a new class of quantum phase transitions driven by tuning of boundary conditions.

\paragraph*{The hard-core Bose--Hubbard model.} The one-dimensional boson Hubbard model~\cite{BH} neglecting nearest-neighbor repulsion is of interest for quantum computation, quantum information, and ultra-cold atoms~\cite{Cold1,Cold2}. Its Hamiltonian consists of three terms. The first, $-\om\sum_n(b_n^\da b_{n+1}+h.c.)$, allows site-hopping of the bosons through the creation and annihilation operators $b_n^\da$ and $b_n$, with $\om$ the hopping matrix element. The second term is $-\mu\sum_nb^\da_nb_n$, where $\mu$ represents the chemical potential of bosons. The third term is $\frac12U(b^\da_nb_n-1)b^\da_nb_n$, where the onsite repulsion $U$ sets the energy scale for the problem. We study in this work a hard-core Bose--Hubbard model having $N$ sites with a single hard-core boson that can also be represented as a spin-$\frac12$ particle. The average particle occupation for each site is $n_0=1/N$ and in the thermodynamical limit $N\rightarrow\infty$ the system approaches a Mott insulator with $n_0=0$~\cite{Mott}. For an open-ended system, the Heisenberg XY model can be solved analytically using a Jordan--Wigner transformation~\cite{JW,Lieb}, which converts a spin-$\frac12$ system into a free spinless fermion chain with nearest-neighbor hopping. In this paper we shall present a hidden quantum phase transition induced by tuning the microscopic coupling strength $J$ between a single pair of nearest-neighbor sites along the closed one-dimensional spin system. This one-dimensional model of Heisenberg XY exchange interaction is shown in Fig.~1.

The Hamiltonian of the total system is written as
\begin{eqnarray}\non
H_{\rm tot}&=&H_0+H(J), \\ \non
H_0&=&\frac{g}{2}\sum_{n=1}^{N}(X_nX_{n+1}+Y_nY_{n+1}), \\
H(J)&=&\frac{J}{2}(X_{s}X_{s+1}+Y_{s}Y_{s+1}), \label{H0}
\end{eqnarray}
where $g$ is the exchange coupling strength between nearest neighbor spins, and $X$ and $Y$ indicate the Pauli matrices $\si^x$ and $\si^y$, respectively. The index $s$ is an arbitrary site number that will be chosen as $s=N$ so that $s+1=1$, with no loss of generality. Units will be chosen so that the coupling strength $g$ is equal to one in the following discussion. The full exciton number is conserved because $[H_{\rm tot}, \sum_nZ_n] = 0$, where $Z_n\equiv\si^z_n$. The quantum phase transition of interest can be observed in the subspace with one exciton. Note that under such conditions the constituents at sites of the system can be replaced by harmonic oscillators (modelling coupled single-mode cavities) or other bosonic modes. Hence the interaction term for two sites can be expressed as $a_n^\da a_{n+1}+a_{n+1}^\da a_n$, where $a$ ($a^\da$) is the annihilation (creation) operator. As for the system investigated in Ref.~\cite{JC}, the present system has a limited number of particles but an infinite number of degrees of freedom.

\paragraph*{Symmetry analysis of the system.} In the language of network topology the model of Eq.~(\ref{H0}) describes a hybrid system consisting of a circle or ring topology and a line segment or bus topology. Figure~1(a) illustrates that all the connection strengths between the nearest-neighbor sites are homogenous except that between the pair $s$ and $s+1$, which is equivalent to the addition of a small line segment described by $H(J)$. When $J=-1$ (in units where $g=1$), the whole system is an open-ended line segment. When $J=0$, the system becomes a circle that is translationally invariant along the sites. The line and the circle are not homeomorphic because the line can be disconnected by removing one site and the circle can not. The circle and the line however are locally homeomorphic only when they have an infinite number of sites. When $J>0$ and $-1<J<0$ the system has no translational invariance.

On the other hand, as $J$ is increased from $-1$ to positive values the system can be regarded as a spin-line segment with increased bending [see Fig.~1(b)]. This effectively changes the boundary condition of the system because of the increasing interaction between the spins at the two ends. This behavior may be expected to occur for any physical system where the site-site interaction is proportional to certain powers of the site separation distance. For example, a Coulomb-like interaction is inversely proportional to the square of the distance between the subsystems and a $10\%$ variation in the site-site distance yields a greater than $25\%$ variation in the mutual coupling strength. Then when the one-dimensional system is folded as shown in Fig.~1(b) until its two ends are near to each other, their mutual interaction is modified sensitively by their separation distance. Thus the model in Eq.~(\ref{H0}) also can be used to investigate the effect of boundary conditions. In a large system with $N$ sites the boundary terms are conventionally expected to be of order $1/N$, and thus to have negligible influence for macroscopic physical quantities. However, we find in this model that microscopic boundary conditions can have a dramatic effect because they can lead to a counterintuitive quantum phase transition.

Henceforth we confine attention to the effect of the control parameter $J$ in Eq.~(\ref{H0}). Roughly the passage from $J=-1$ to $J=0$ is expected to be connected closely to the transition from one well-defined topology to another well-defined topology, while the passage from $J=0$ to $J>0$ corresponds to a transition from a well-defined topology to a hybrid one. As will now be demonstrated, we observe a hidden quantum phase transition around $J=0$ in this system that is reflected in the singularity properties of the ground and the first-excited states.

\paragraph*{Ground state properties.} First, let us consider the dependence of the ground state energy on system size, which is a functional of the boundary condition parameterized by $J$ or the extra strength between two selected neighbor sites $1$ and $N$. We display the two lowest derivatives of the ground state energy $E_0$ with respect to $J\in[-1,1]$ in Figs.~2(a)--2(d) for systems with $N=50, 100, 1000,$ and $10,000$ sites, respectively. Neither derivative changes significantly until $J$ approaches zero. The first derivative $\pa_JE_0$ exhibits a rapid decrease with increasing $J$ after passing through $J=0$, while the second derivative $\pa^2_{J^2}E_0$ exhibits a cusp at $J=0$, with both behaviors becoming more sharply defined as the site number is increased. Thus, the variation of the energy derivatives indicates that the model experiences a quantum phase transition at the critical point $J=0$ that is controlled entirely by tuning the connection strength between a single pair of spins on the boundary.

A more sensitive indicator of the quantum phase transition is afforded by changes of the wavefunction near the critical point. Figure~3 displays the calculated overlap of the ground state wavefunction $\Psi_J$ with the limiting wavefunctions $\Psi_{-1}\equiv\Psi_{J=-1}$ and $\Psi_{0}\equiv\Psi_{J=0}$. The ground state wavefunction overlaps exactly the initial state $\Psi_{-1}$ as $J$ is increased in the interval $J\in[-1,0)$, and then transforms rapidly into $\Psi_0$ near the single point $J=0$, with the transition becoming increasingly sharp with larger $N$. When $N$ is increased to $10,000$, the overlap exhibits a critical behavior through its dependence upon $J$, the corresponding transition becomes essentially a step function, and the derivative $\partial_J\langle\psi_J|\psi_{-1}\rangle$ shown in the top-right inset in Fig.~3 forms an increasingly clear cusp as $N$ increases. The finite-size scaling of $\partial_J\langle\psi_J|\psi_{-1}\rangle$ is shown in the lower-left inset in Fig.~3, where the critical exponent $\nu$ is found numerically to be around $0.63$. Even more interesting, after $J$ increases through the critical point $J=0$ the ground state wavefunction jumps immediately to another state $\Psi_{J>0}$, that is fully orthogonal to $\Psi_{-1}$ (their overlap vanishes) and nearly orthogonal to $\Psi_{J=0}$ (their overlap is also very close to zero). Thus Fig.~3 indicates unequivocally a quantum phase transition at $J=0$ involving a level crossing in which the ground state is replaced by a new ground state essentially orthogonal to the original state. Combining these results with the results for the derivatives of the ground state energy in Fig.~2, this new kind of quantum phase transition can be categorized as second-order.

This is a remarkable result in that this new kind of quantum phase transition may be attributed purely to a microscopic boundary condition of the quantum system, without invoking any applied external or internal field. According to the Hellmann$-$Feynman theorem~\cite{HF}, $\pa_JE_0=\la\Psi_J|\pa_JH_{\rm tot}|\Psi_J\ra=1/2\la\Psi_J|(X_1X_N+Y_1Y_N)|\Psi_J\ra$, where $|\Psi_J\ra$ denotes the ground wavefunction of the system with special $J$. Thus the expectation value of the boundary interaction term $X_1X_N+Y_1Y_N$ may be taken as the order parameter associated with the quantum phase transition. The dimensionless interaction strength $J$ between a single pair of sites along the circle, which is a quantity that could be modified easily in experiments, previously was thought to be unimportant as $N \rightarrow \infty$.

Therefore one can see that the transition of the system from a chain to a circle as $J$ varies from $J=1$ (a mixture of line-segment and circle topologies) to $J=0$ (a perfect circle) {\em causes no phase transition} due to the invariance of the ground state. However, the transition of the {\em boundary condition} from a state with translation symmetry ($J=0$) to a state of broken-symmetry ($J>0$) leads to a clear quantum phase transition due to a two-step sudden change in the ground state. After the critical-point, the ground state becomes $\Psi_{-1}^{\bot}$ for $J>0$, satisfying $\la\Psi_{-1}^{\bot}|\Psi_{-1}\ra=0$.

\paragraph*{Discussion and Conclusion.} The microscopic details of the individual building blocks of a system generally are believed to not be crucial to quantum phase transitions. Rather, the collective behaviour is thought to be controlled by general properties of the interaction between the building blocks (spins or harmonic oscillators), and the quantum phase transition is induced by tuning an ``external'' parameter (field strength, pressure, doping concentration, \ldots). We have shown here a new class of quantum phase transition induced by a microscopic boundary condition as well as translational symmetry in a finite-size, one-dimensional spin model with no external applied field that defies this common knowledge. The quantum phase transition exhibited here does not require a noncommuting condition between different constituents of the Hamiltonian. At the critical point, we demonstrate a nonanalytic behavior of the second-order derivative of the ground eigen-energy and a sudden change of the ground state wavefunction. This special quantum phase transition might also be understood by concepts developed for symmetry-protected topological phases except that the system analyzed here has a {\em local order parameter} that can characterize the phase transition. Our result shows that variation of a single microscopic link between two constituents (which could be regarded as a topological defect) can produce a sudden change in the ground-state properties of a system with large size. Therefore we have exhibited a new kind of quantum phase transition that is controlled by tuning a microscopic boundary condition rather than a macroscopic field that could have large practical implications for applications in quantum information, quantum control, and quantum computing.

\paragraph*{Acknowledgements.} We thank Dr. Z. M. Wang for his early participation.

\clearpage

\begin{figure}[htbp]
\centering
\includegraphics[width=0.8\linewidth]{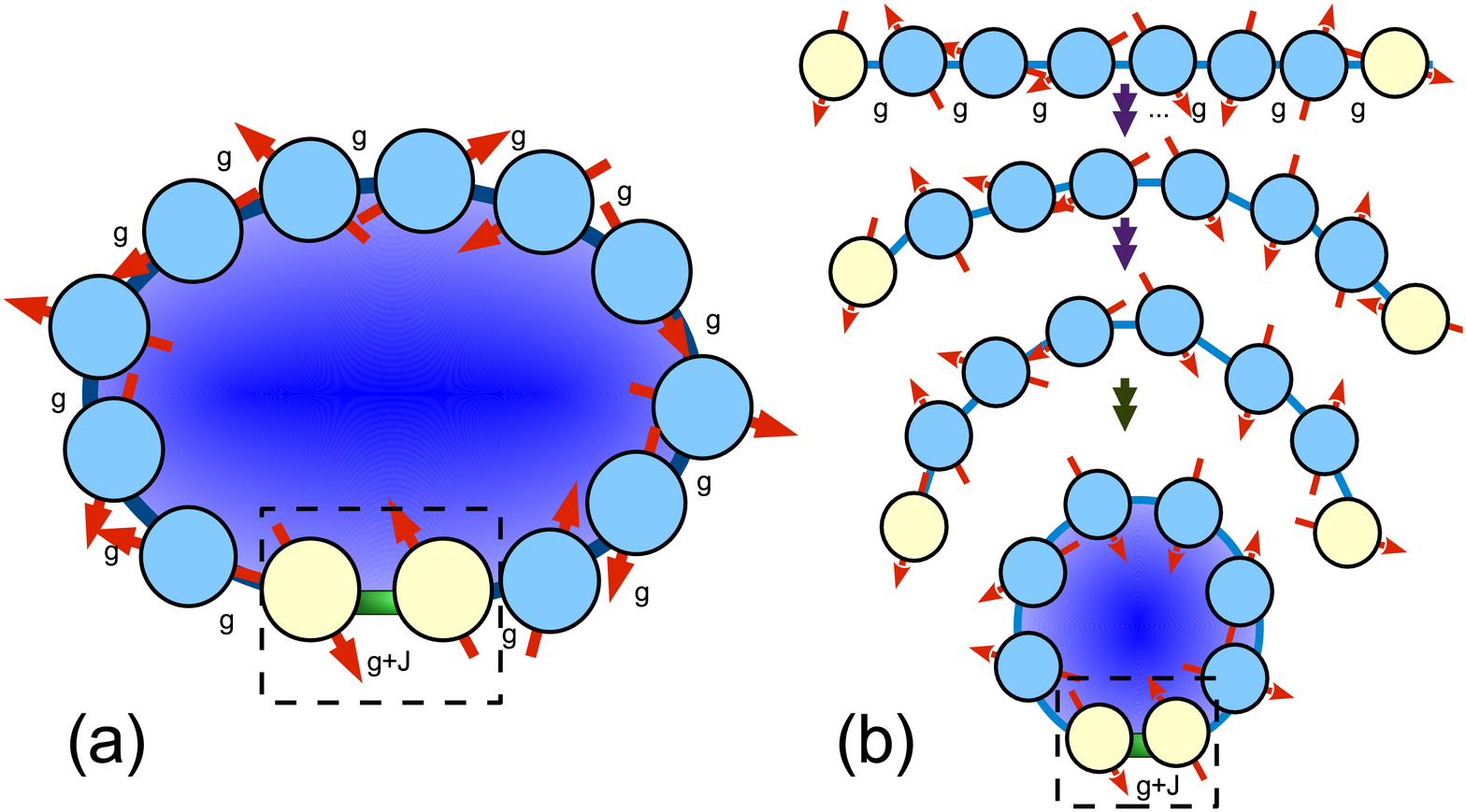}
\caption{(Color online) A model system that demonstrates a new kind of quantum phase transition. (a)~This one-dimensional spin system with a finite number of sites $N$ can be viewed as the sum of two topologies: a spin-circle with isotropic interaction among nearest neighbor sites and a spin-line segment consisted of two sites with arbitrary coupling strength. (b)~With increased or decreased bending the coupling between the two ends of the system becomes stronger or weaker.}\label{sketch}
\end{figure}

\clearpage
\begin{figure}[htbp]
\centering
\includegraphics[width=0.9\linewidth]{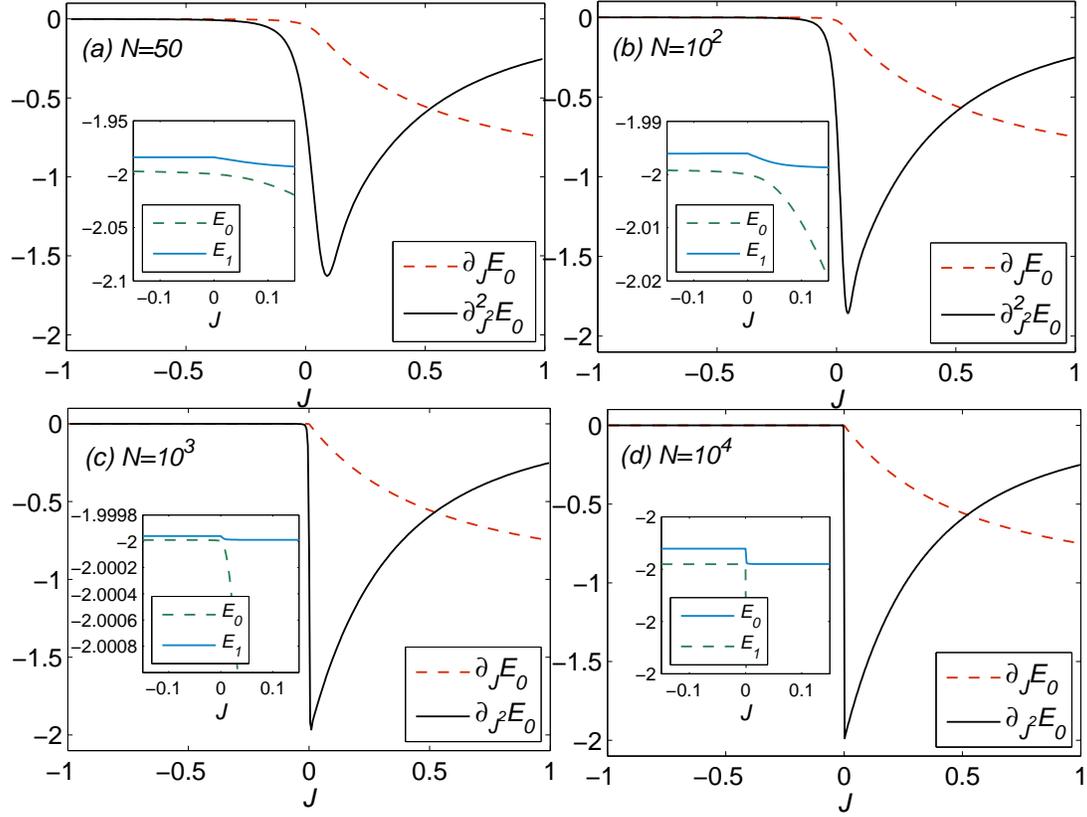}
\caption{(Color online) The first derivative of the ground state energy with respect to $J$ (dashed red curve) and the second derivative (solid black curve) as functions of $J$. The sizes of the system are (a) $N=50$, (b) $N=10^2$, (c) $N=10^3$, (d) $N=10^4$, respectively.}\label{EDE}
\end{figure}

\clearpage
\begin{figure}[htbp]
\centering
\includegraphics[width=0.9\linewidth]{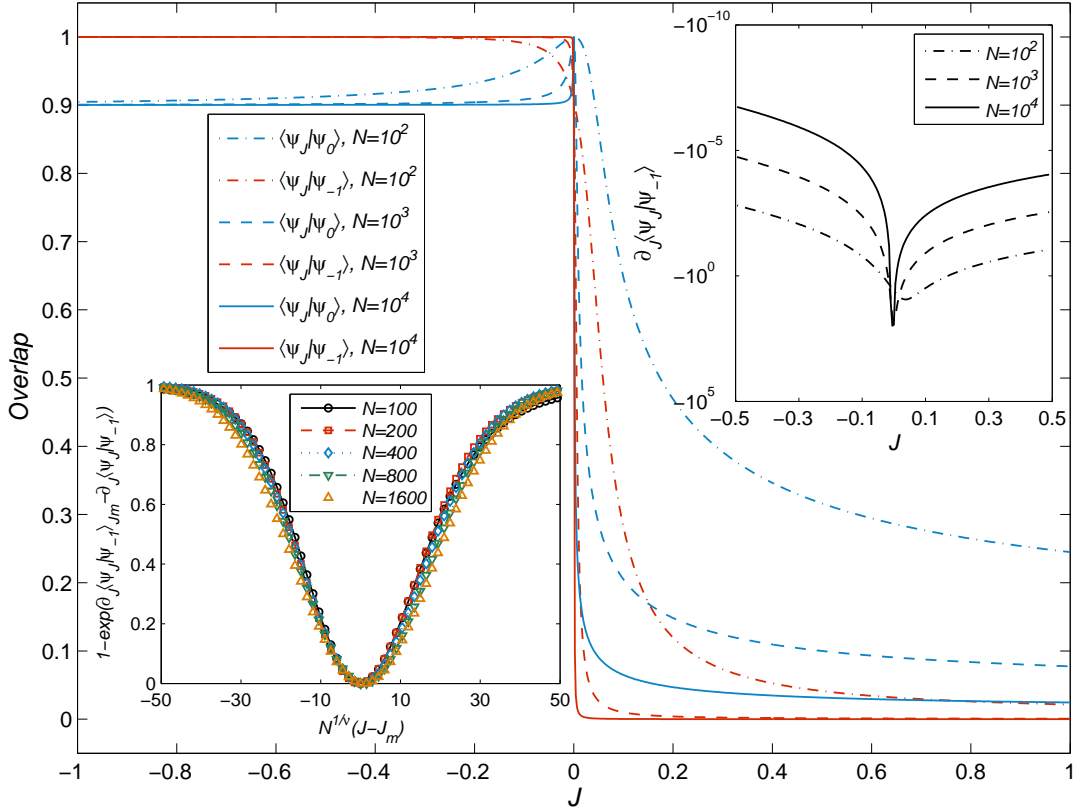}
\caption{(Color online) The overlap between the ground state $\Psi_J$ within the interval $J\in[-1,1]$ and the special case with $\Psi_{-1} \equiv \Psi_{J=-1}$ (red lines) and that with $\Psi_0 \equiv \Psi_{J=0}$ (blue-lines). Different sizes of the system are represented by different type of lines (dot-dashed lines for $N=10^2$, dashed lines for $N=10^3$, and solid lines for $N=10^4$). Top-right inset: The first derivative of the overlap $\langle\psi_J|\psi_{-1}\rangle$ with respect to $J$ for different system sizes. Lower-left inset: Finite-size scaling of the derivative $\partial_J\langle\psi_J|\psi_{-1}\rangle$ with the number of sites $N$. The derivative $\partial_J\langle\psi_J|\psi_{-1}\rangle$ is a function of $N^{1/\nu}(J-J_m)$ with a critical exponent $\nu\approx0.63$, and $J_m$ is the position of the minimum of $\langle\psi_J|\psi_{-1}\rangle$. All the data from $N=100$ to $N=1600$ collapse onto a single curve.}\label{Overlap}
\end{figure}

\end{document}